# PSA: The Packet Scheduling Algorithm for Wireless Sensor Networks

[1]C. Jandaeng,[1]W. Suntiamontut, [2]N. Elz

[1]Centre of Excellence in Wireless Sensor Networks,
Department of Computer Engineering, Prince of Songkla University
`cjundang@gmail.com`

[2]Computer System and Network Laboratory,
Department of Computer Science, Prince of Songkla University,
Hat Yai, Thailand 90112

## Abstract

*The main cause of wasted energy consumption in wireless sensor networks is packet collision. The packet scheduling algorithm is therefore introduced to solve this problem. Some packet scheduling algorithms can also influence and delay the data transmitting in the real-time wireless sensor networks. This paper presents the packet scheduling algorithm (PSA) in order to reduce the packet congestion in MAC layer leading to reduce the overall of packet collision in the system The PSA is compared with the simple CSMA/CA and other approaches using network topology benchmarks in mathematical method. The performances of our PSA are better than the standard (CSMA/CA). The PSA produces better throughput than other algorithms. On other hand, the average delay of PSA is higher than previous works. However, the PSA utilizes the channel better than all algorithms.*

## Keywords

*packet collision, packet scheduling algorithm, wireless sensor networks*

## 1. INTRODUCTION

A wireless sensor network is a self-configured network containing numerous small sensor nodes. Each node consists of sensing modules, a processing unit, radio frequency components and power sources [1]. They organize and communicate among themselves in an ad-hoc fashion. The wireless sensor network technology has been deployed in several applications such as health care monitoring systems, home automation and environment monitoring systems [2]. These applications require inexpensive facilities and little manual maintenance. According to the application requirements, each node has been implemented using a low-power microcontroller and radio module. In addition, each node is supplied with a small battery. Energy usage is the indicator of network lifetime [3].

All sensor nodes share a single communications channel using a multiple access protocol. The packet transmission may lead to a time overlap of two or more packet receptions, called collisions. The packet collision problem causes packet loss, packet retransmission, decreasing throughput, increased delay/latency and increased wasted energy consumption. Many research works on the MAC protocol have been proposed to solve the packet collision problem [4] such as Spatial TDMA traffic-adaptive medium access protocol (TRAMA) [5], Sensor MAC (SMAC) [6], and Timeout MAC (TMAC) [7]. A MAC protocol based on IEEE 802.15.4 was developed for low-power communication. The IEEE 802.15.4 MAC protocol uses a random back off in order to reserve and access the channel. A node is authorized to send the packet when the channel is idle. In contrast, random back off is activated when the channel is busy. Unfortunately, this technique will not work properly when used in a large scale wireless sensor network.

Time Division Multiple Access (TDMA) is a solution to reduce the packet collision problem. Total transmission time is divided into frames and each frame is divided into time slots. After that each time slot will be assigned to a sensor node to guarantee that every node is granted permission to send a packet in its time slot guaranteeing collision avoidance. Latency directly varies with frame length. On other hand, throughput inversely varies with frame length. There have been many approaches presented to minimize the frame length and maximize the throughput which are explained in section 2.

All previous works illustrated above are proposed for an ad hoc network. All devices are powerful nodes having unlimited energy. In contrast, sensor nodes are resource constrained having limited energy and low processing power. Therefore, the characteristics of the scheduling algorithm for a sensor network should be simplicity and efficiency. This paper therefore proposes a new algorithm based on the greedy technique that is simple and easy to implement in resource constrained devices. This paper will explain the proposed PSA and describe the evaluated results of the performance using mathematical results.

The remainder of the paper is organized as follows. First we briefly explain the packet collision problem and previous works in section 2. After that, the packet scheduling algorithm is described in detail in section 3. The performance comparisons using the mathematical results are presented in section 4. Finally we give the conclusion about the performance of the proposed packet scheduling algorithm in section 5.

## 2. PREVIOUS WORKS

Y. Peng *et al* [8] presented the TDMA with a scheduling matrix. The row of the matrix denotes frame length while the column of the matrix denotes nodes. The members of the matrix represent transmission authorization. In [8], they proposed to optimize the number of rows that refers to the frame length with Tabu search and greedy algorithm. This approach can reduce the average latency and produce high throughput in a dense area.

G. Wang and N. Ansari [9] have proved that the scheduling matrix optimization is an NP-complete problem. They also proposed an approximation method, mean field anneal (MFA) to optimize the schedule matrix. The matrix optimization is divided into two phases: minimize frame length and maximize throughput. More recently approximation methods have been proposed. S. Salcedo-Sanz *et al* [10] minimized frame length with a neural network (NN) and maximized throughput with a genetic algorithm (GA), whereas J. Yeo *et al* [11] applied the sequence vertex coloring (SVC) in both phases. S. Haixiang and W. Lipo [12] proposed a hybrid algorithm which combined back tracking sequential coloring (BSC) and noisy chaotic neural network (NCNN) to optimize the scheduling matrix. BSC-NCNN gives the minimal average time delay, while the NN-GA provides higher throughput.

I. Ahmad *et al* [13, 14]. proposed an idea to avoid packet collision. The network topology is represented by a finite state machine (FSM). The set of nodes are grouped with the maximal compatibles and incompatibles concept. This method begins by setting up a number of groups that equals the number of nodes. After that, combine groups together under the condition that no nodes in the same group are neighbor nodes. Finally, all sensor nodes are grouped in many groups and they can send packet in the same time without collision. The number of groups is frame delay while the summation of number of node in all groups is throughput. This idea leads to minimize latency and maximize throughput.

## 3. THE PACKET SCHEDULING ALGORITHM

The Packet Scheduling Algorithm (PSA) is the algorithm that schedules all packets from application layer and network layer in order to reduce network congestion in the data link layer to avoid the packet collision. When the PSA is implemented, packet collisions will be minimized with increasing of throughput as a by product. A PSA based on a greedy algorithm is a simple algorithm and easily implemented in a sensor node. The basic assumptions of the PSA are defined as follows. All packets communicate via IEEE 802.15.4 standard [15] that avoids packet collision with a simple CSMA/CA mechanism. All sensor nodes must know the information of at least 2-hops neighbor nodes. Finally, time synchronization, neighbor discovery, and routing protocols are not considered in this work.

### 3.1 Definitions

The node color represents *node status*. Two functions, *combine*() and *match*() are used to reduce the frame length. The node statuses and their functions are defined below:

**Definition 1:** *Node status* is represented by a node color for each time slot. A black node can send any packet with a guarantee of no collision. If a white node requests to send a packet, its packet may collide. A gray node is in the initial status with no guarantee with regard to packet collision. Finally, a gray node can change status to the other colors with *combine*() and *match*() functions. Figure 1 shows an example of 15 nodes status. The color of each node is set corresponding to its status which could be either black or gray (with "x") or white.

| S | 1 | 2 | 3 | 4 | 5 | 6 | 7 | 8 | 9 | 10 | 11 | 12 | 13 | 14 | 15 |
|---|---|---|---|---|---|---|---|---|---|----|----|----|----|----|----|
| 1 |   | x |   | x |   |   | ■ | x | x | x |   |   | x | x |   |

Figure 1: Node status in PSA algorithm

**Definition 2:** The function *combine*() is used to reduce the frame length by combining two frames. The two frames must be tested with the *match*() function before the combination. The will be merged if the *match()* function returns valid. In the combination process, the status of a node can be changed to another color as defined below. Let $A$, $B$ and $R$ denote the frame and $A_i$, $B_i$, and $R_i$ are node status in the $i^{th}$ time slot of $A$, $B$, and $R$; $V$ denotes the set of nodes and $R=combine(A,B)$.

$$\forall i_{i\in V}, R_i = \begin{cases} A_i & \text{if } B_i \text{ is GRAY} \\ B_i & \text{otherwise.} \end{cases}$$

The above equation also means that:

2.1) the node status of frame $R$ can be replaced with the status of $A_i$ if $B_i$ is a gray node.

2.2) Otherwise, it will be replaced with $B_i$.

2.3) *combine*(A, B)=*combine*(B, A) if *mach(A, B)*.

| S | 1 | 2 | 3 | 4 | 5 | 6 | 7 | 8 | 9 | 10 | 11 | 12 | 13 | 14 | 15 |
|---|---|---|---|---|---|---|---|---|---|----|----|----|----|----|----|
| A | x | x |   | x |   |   |   | x | x |   |   | ■ | x | x |   |
| B | ■ |   |   |   |   |   |   | x | x | x | x | x | x | x | x |

| R | ■ |   |   |   |   |   |   | x | x |   |   | ■ | x | x |   |

Figure 2: A Result of *combine*() function in PSA algorithm

Figure 2 shows a result of *combine*() function. The outcomes of the definition 2.1 are $R_8 - R_{15}$ and $R_1$-$R_7$ come from the definition 2.2. From definition 2.1, the gray nodes can be changed to black or white because the gray node is an unknown status.

**Definition 3:** The *match*() function is used to validate two frames before combination. Only two matched frames can be combined. The notation *match*(*A*,*B*) means that the frame *A* and *B* are matched before the combination process in definition 2 starts. Frame *A* and *B* are matched only if all nodes in these two frames meet this condition:

$$match(A, B) \leftrightarrow \forall i_{i \in N} A_i = GRAY$$

$$\lor (A_i = BLACK \land B_i = GRAY)$$

$$\lor (A_i = WHITE \land B_i \neq BLACK)$$

The condition is explained that:

3.1) $A_i$ is gray node while $B_i$ is any status because $A_i$ is unknown status and can be replaced with any status of $B_i$.

3.2) $A_i$ is black node and $B_i$ is gray node mean that $A_i$ is reserved for node $i^{th}$. They can be combined because $B_i$ can be changed to any status.

3.3) $A_i$ is white node while $B_i$ is not black node. If any node is blocked in frame $A$, the same node in frame $B$ must be blocked or still as unknown status.

| S | 1 | 2 | 3 | 4 | 5 | 6 | 7 | 8 | 9 | 10 | 11 | 12 | 13 | 14 | 15 |
|---|---|---|---|---|---|---|---|---|---|----|----|----|----|----|----|
| A |   | x |   | x |   |   | ■ | x | x | x  |    |    | x  | x  |    |
| B | x | x |   | x |   |   |   | x | x |    |    | ■  | x  | x  |    |
| C | ■ |   |   |   |   |   |   | x | x | x  | x  | x  | x  | x  | x  |

Figure 3: Result of *match*() in PSA algorithm
(Frame A and B are matched while Frame A and C are not matched)

The figure 3 is and example of the *match*() function. Frame $B$ matches with frame $C$ while frame $A$ does not match with frame $B$. When we determine slot $B_1$, $B_2$, $B_4$, and $C_8$-$C_{15}$, we found that they match because of definition 3.1. Slot $B_3$ and $B_5$-$B_7$ match because all nodes are white nodes in $B$ and $C$ as shown in definition 3.3. In the same way, slot $C_1$ and $B_{12}$ also match because of definition 3.2. There are black whereas the other time slots are gray nodes. From *match*(*A*,*B*), we can conclude that they do not mach because $A_7$ and $B_7$ conflict with definition 3.2. One of them is black while the other is white. Thus, they could not be combined.

## 3.2 Algorithm

The wireless sensor network is represented based on a undirected graph $G$=($V,E$) where $V$ represents the set of sensor nodes and $E$ represents the set of edges. In the case of ($u,v$)∈$E$, it means that node $u$ sends packets directly to node $v$, they are one hop apart. Furthermore, if $u$ and $v$ are not one hop apart but have an intermediate node $k$ such that ($u, k$)∈$E$ and ($k, v$)∈$E$, nodes $u$ and $v$ are said to be two hops apart.

This algorithm consists of three phases. First, the network topology represented in $G$=($V,E$) is transformed to scheduling matrix, $S$, called *scheduling matrix initiation* phase. After that we reduce the frame length of scheduling matrix with *frame length minimization* phase in order to minimize the average delay. The final phase is to maximize the throughput and channel utilization that called *throughput maximization* phase. The details of all phases are explained below:

**Phase I) Scheduling matrix initiation**

The scheduling matrix initiation is the first phase. The network topology is represented in $V$ denotes the set of sensor nodes, and $E$ which denotes the set of edges. Both $V$ and $E$ are the input of algorithm 1 and the scheduling matrix, $S$, is the result of this phase. The square scheduling matrix consists of columns and rows sized |$V$|. Each row is a list of time slots called frame, $F_n$.

The $f_{ni}$ is the status of node $i$ in frame $n$ and is represented by a color as explained before. Therefore, the number of rows in the scheduling matrix is called frame length.

---

**Algorithm 1** scheduling matrix initiation

---

1:  **for** $u \in V$ **do**
2:      Set GRAY to all member for list, $F_u$
3:      $f_{uu} = BLACK$
4:      **for** $v \in V$ **do**
5:          **if** $(u, v) \in E$ **then**
6:              $f_{uv} = WHITE$
7:              **for** $k \in V$ **do**
8:                  **if** $(k, v) \in E$ **then**
9:                      $f_{uk} = WHITE$
10:                 **end if**
11:             **end for**
12:         **end if**
13:     **end for**
14:     $S = S \cup \{F_u\}$
15: **end for**

---

Algorithm 1 is explained that all node statuses in frame, $F_u$, are set to gray. The node, $f_{uu}$, is set to black mean that this frame is granted for node $u$. All adjacency nodes, $(u, v) \in E$, are set to white in order to prevent direct collision and all adjacency nodes, $(k, v) \in E$, are set to white in order to prevent hidden collision. Finally, frame, $F_u$, is added to the schedule matrix, $S$. This algorithm will be repeated for every sensor node in $V$. We get the scheduling matrix, $S$, and frame length $|V|$ when the first algorithm finishes.

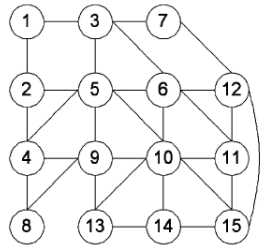

(a) 15 node topology

(b) scheduling matrix initiation

(c) frame length minimization

(d) throughput maximization

Figure 4: The PSA algorithm

Figure 4a and 4b give an example of algorithm in the first phase. A network topology with 15 nodes is shown in figure 4a. The scheduling matrix in figure ab is the result from phase 1. This matrix sized 15$x$15 consists of 15 black slots that are granted as one slot for each node. Moreover, there are 70 gray slots that can be changed with the next phases. The frame length can be optimized with algorithm 2, while the gray slots are changed to black or white using algorithm 3.

**Phase 2) Frame length minimization**

The frame length indicates the average waiting time of a sensor node. For example, the node 1 must wait for 14 frames in order to send a packet in its next turn. To minimize the frame length of the schedule matrix, we group all frames with *combine*() and *match*() functions as defined in the previous section based on the greedy algorithm.

The input of this phase is the scheduling matrix, *S*, while the output is the minimized frame length of scheduling matrix. Let $F_a$, $F_b$, and $R$ denote frames in the scheduling matrix. The algorithm of phase 2 is explained below whereas the *max()* function is defined in algorithm 3 .

---

**Algorithm 2** frame length minimization

1:  **loop**
2:    **if** $F_a = max(S, NULL)$ **and** $F_b = max(S, F_a)$ **and** $match(F_a, F_b)$ **then**
3:      $R=combine(F_a, F_b)$
4:      $S = S - \{F_a, F_b\}$
5:      $S = S \cup \{R\}$
6:    **else**
7:      **return** $S$
8:    **end if**
9:  **end for**

---

The *max( )* function finds the frame of S with the maximum number of gray nodes (other than one already chosen frame).

---

**Algorithm 3** *max(S, F)* function

       **Input :** *S* is scheduling matrix and *F* is frame
1:    $R = NULL$
2:    $g = 0$
3:    **for** $r \in S$
4:      **if** $r == F$ **then**
5:        **continue**
6:      **if** $gray(r) > g$ **then**
7:        $g = gray(r)$
8:        $R = r$
9:      **end if**
10:   **end for**
11:   **return R**

---

The weighting function is shown in the second line of algorithm 2. The algorithm selects two frames that contain the maximal gray slot because they have a high probability of matching successfully and provide the most gray slot after combination.

The algorithm repeats all statements until there are no matched frames according to the condition in the second line. For each round, it finds two frames from the schedule matrix under two conditions: 1) They are the first and second frames that provided the maximum gray slot and 2) two frames must follow the definition 3. After that, the selected frames are removed and combined to be the new frame, $R$. The new frame, $R$, is added into the schedule matrix. If the condition in the seventh line is true, this phase will stop and return an optimal scheduling matrix, $S$. Finally, we get the new schedule matrix that provides a minimal frame length as shown in figure 4c.

The numbered black slots from this phase are equal to the initial scheduling matrix. However the frame length and number of gray slots are reduced. The next phase replaces gray slots with black slot in order to increase throughput. Phase 3 still relies on *match*() and *combine*().

**Phase 3) Throughput maximization**

Throughput maximization is the last phase of PSA. This phase increases the number of black nodes by replacing gray with black or white color in order to increase the throughput. However, the node replacement must follow *match*(), *combine*() and algorithm 4. The input of this phase is the scheduling matrix shown in figure 4c. The algorithm eliminates gray slots and replace with black or white. Moreover, the initial scheduling matrix, $iS$, produced by the first phase is used in this phase. At the end of this phase, the new scheduling matrix, $S$, is composed entirely of black and white slots.

| **Algorithm 4** throughput maximization |
| --- |
| 1:   **for** $u \in V$ **do** |
| 2:       **for** $F_v \in iS$ **do** |
| 3:           **If** $f_{vu} = GRAY$ **and** $match(iF_u, F_v)$ **then** |
| 4:               $F_v = combine(iF_u, F_v)$ |
| 5:           **end if** |
| 6:       **end for** |
| 7:   **end for** |
| 8:   replace all gray nodes with white nodes |

The main idea of this phase is to replace all gray slot that are valid with *match*(). The scheduling matrix is traversed in column order to find a gray slot. Fore example in figure 4c, the first node contains four gray slots and one black slot. The second frame of the first node is a gray slot. That means the first node may transmit the packet without collision. In order to ensure that the first node can send packet in this frame, the frame $iS_1$ and $F_2$ are tested with *match*() function. They are merged with *combine* function only if they are matched. After frame combination, the gray slot of the second node in second frame is replaced with white slot because of the *combine*() function. All gray slots in the first column are replaced with black that result in gray slots of the other columns are changed to be white slot. The fourth column is changed to white slot. Therefore, the eighth column will be processed in the next step. Finally, a optimal scheduling matrix is generated and shown in figure 4d.

The packet scheduling algorithm transforms the network topology to be a scheduling matrix. All node members in the matrix are set to black, gray or white color. The PSA combines two frames that tested by *match*() and *combine*() functions in order to reduce the frame length and increase black slots. Both frame length minimization and throughput maximization phases are based on greedy algorithm. A mathematical evaluation by comparing with the previous works in terms of throughput, average delay and channel utilization will be presented in the next section.

## 4. MATHEMATICAL EVALUATION

Packet collision minimization is the primary goal of the proposed algorithms in the broadcast scheduling problem (BSP). However, the packet scheduling cause effects upon network such as average delay, throughput, and channel utilization. This section explains the three performance metrics that are used to evaluate the proposed algorithm and compare the PSA with the previous algorithms using network benchmarks.

## 4.1 Performance Metrics

There are three performance metrics for mathematical evaluation of the PSA algorithm which are throughput, average delay, and channel utilization.

**Throughput (σ: slots)** It is the number of reserved time slots, or black slots, that are assigned to sensor node. The throughput is calculated using the equation below. The schedule matrix, *S*, is of size |*V*|x|*S*|. |*V*| denotes the number of nodes and |*S*| denotes the frame length, and $s_{ij}$ is the status of node in each time slot.

$$\sigma = \sum_{i=1}^{N} \sum_{j=1}^{L} s_{ij}$$

when

$$s_{ij} = \begin{cases} 1 & \text{if } s_{ij} \text{ is black} \\ 0 & \textit{otherwise.} \end{cases}$$

**Averaged delay (τ: frames)**. This indicates the waiting time of a sensor node between opportunities to transmit. The average delay is calculated by the equation below. This metric depends on the frame length and number of black slots per node. If any algorithm can reduce the frame length and generate the same throughput, the average delay will different. The distribution of black slots can determine the average delay. A high distribution gives a lower average delay compared to a low distribution.

$$\tau = \frac{|S|}{|V|} \sum_{i=1}^{|V|} \left( \frac{1}{\sum_{j=1}^{|S|} s_{ij}} \right)$$

**Channel Utilization (η :%):** We trade-off between the throughput and the average frame delay. More frames mean more available time. On the other hand, a high frame length can increase the averaged frame delay. Therefore, the channel utilization is the best metric to measure the performance of the algorithm. The channel utilization is calculated from the equation below.

$$\eta = \frac{\sigma}{|S|x|V|} x100$$

## 4.2 Results and Discussions

This section compares the PSA with other algorithms. All algorithms are tested with the network topology introduced by [9], which has become the benchmark test cases for the broadcast scheduling problem. The network benchmarks consist of three topologies with 15, 30, and 40 nodes as shown in figure 5. The maximum of neighbor node of all benchmarks are 7 nodes as indicated by the minimal frame length of the scheduling matrix.

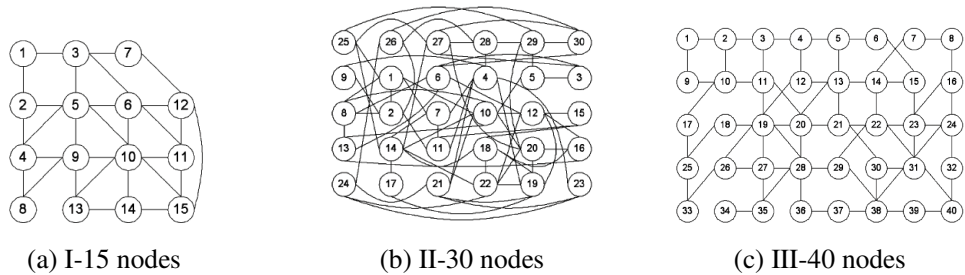

| (a) I-15 nodes | (b) II-30 nodes | (c) III-40 nodes |

Figure 5: Network benchmarks

All benchmarks are scheduled with the PSA and the other algorithms. The scheduling matrixes of
PSA are shown in figure 6. Each matrix consists of frame (row) and sensor node (column). The
frame lengths of the three benchmarks are 10, 14, and 11, and throughput (black slots) are 26, 53,
94 slots. Each frame consists of time slots that are filled with black or white color. Node $j$ in
frame $i$ filled with black color means that node $j$ sends a packet in frame $i$ with no collision. For
example the first frame in figure 6a is reserved for node 3, 8, and 14. Thus, node 3, 8, and 14 are
granted permission to send a packet in this frame while the other nodes are blocked.

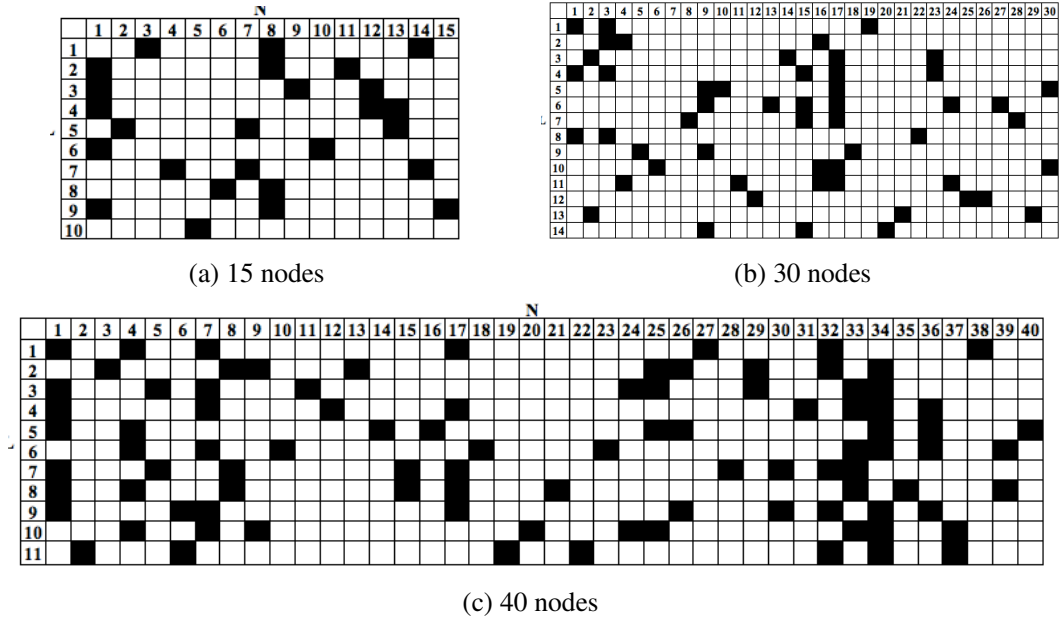

(a) 15 nodes          (b) 30 nodes

(c) 40 nodes

Figure 6: scheduling matrix

The performances of all algorithms are shown in table 1. The performance metrics of three
scheduling matrixes are calculated with the equations in section 4.1 and compared with the other
algorithms. We compare the PSA with the previous works using the statistical method: one
sample t-test. The PSA is compared with the average of old methods for each performance
metrics and topology. The hypothesis assumption is the performance of PSA differs from the
previous works. We found that the performance metrics are mostly different from the previous
works with the confidence level at 95% in contrast with the channel utilization of 40 nodes
topology.

Table 1: Performance comparison

| benchmark | | TABU | HNN | BSC | MFA | SVC | FSM | PSA |
|---|---|---|---|---|---|---|---|---|
| 15 nodes | $\|S\|$ | - | - | 8 | 8 | 8 | 8 | 10 |
| | $\sigma$ | 20 | - | 20 | 18 | 18 | 20 | 26 |
| | $\tau$ | - | 6.80 | 7.00 | 7.20 | 7.20 | 6.84 | 7.63 |
| | $\eta$ | - | - | 16.67 | 15.00 | 15.00 | 16.67 | 17.33 |
| 30 nodes | $\|S\|$ | - | - | 10 | 9 | 11 | 10 | 14 |
| | $\sigma$ | 37 | - | 35 | 38 | 37 | 35 | 53 |
| | $\tau$ | - | 9.20 | 9.30 | 10.67 | 9.99 | 9.20 | 10.99 |
| | $\eta$ | - | - | 11.67 | 10.56 | 11.21 | 11.67 | 12.62 |
| 40 nodes | $\|S\|$ | - | - | 8 | 8 | 8 | 8 | 11 |
| | $\sigma$ | 68 | - | 77 | 71 | 60 | 64 | 94 |
| | $\tau$ | - | 5.80 | 6.30 | 6.99 | 6.76 | 6.00 | 8.39 |
| | $\eta$ | - | - | 24.06 | 19.72 | 18.75 | 20.00 | 21.36 |

The TABU focused on throughput maximization while the HNN focused on average delay minimization. Therefore, they show only throughput or average delay while other approaches determined both average delay and throughput concurrently.

Most algorithms reduce the frame length to 8.0 frames on benchmark I and III. The average frame length of benchmark II is $10.0 \pm 1.29$ frames (average value $\pm$ 95%CI). The PSA reduces the frame length significantly less than other algorithms. The frame length of PSA in the three benchmarks is more than previous works by 25%, 40%, and 37.5% respectively. The average throughputs for all benchmarks are 19.2, 36.4 and 68 slots, respectively. There is 95% confidence to believe that throughput of each algorithm is not different. The PSA generates the free time slot (black node) significantly more than previous works up to 30.00%, 39.47%, and 22.07% on 15, 30, and 40 nodes respectively.

The average delay ($\tau$) and channel utilization ($\eta$) are calculated from the equation in section 3.1. The average delay varies directly with frame length and throughput whereas channel utilization also varies directly with throughput but varies indirectly with frame length. The average delays of PSA are more than the other algorithms. The average delays of previous works are 6.96, 9.67, and 6.37 for the three network benchmarks. The delays of each algorithm do not difference significantly but results from PSA are greater than all other algorithms. Because PSA has a frame length longer than the other algorithms, this disadvantage causes an advantage in free slot allocation and leads to throughput increasing. The PSA generates significantly more throughput than other algorithms because there is more free space in the scheduling matrix. Because of the maximal throughput, the channel utilization of PSA is better than most algorithms and most benchmarks except the BSC in 40 nodes topology.

Table 2 shows the first and second algorithms that produce the lowest average delay, the highest throughput, and the highest channel utilization. There are three algorithms that have better performance than other algorithms such as PSA, FSM, and HNN.

Table 2: The first and second algorithm ordered by performance metrics

| benchmark | | $\tau$ | $\sigma$ | $\eta$ |
|---|---|---|---|---|
| (1) | 1st | HNN | PSA | PSA |
| 15 nodes | 2nd | FSM | TABU, FSM, HNN | BSC,FSM |
| (2) | 1st | FSM, HNN | PSA | PSA |
| 30 nodes | 2nd | BSC | MFA | BSA, FSM |
| (3) | 1st | HNN | PSA | BSC |
| 40 nodes | 2nd | FSM | BSC | PSA |

To compare average delay, HNN is the algorithm that reduces the packet collision under the minimum average delay and FSM is the second. The average delay of PSA is more than other methods because it has the highest frame length. Although the PSA generates the highest throughput, it is not enough to minimize the average delay. Throughput and frame length are not the main factors that affect the average delay. The number of slots per node in the scheduling matrix is the main factor instead. If each node has been allocated fairly, it will result in lower average delay. Figure 6a is the example. The PSA allocates 5 slots for node 1 while most other nodes are allocated only 1 or 2 slots. In contrast, the FSM gives approximately the same number of allocated slots for all nodes. Because of this, the average delays of FSM are less than PSA in spite the throughput of PSA being more than FSM.

The PSA utilizes the channel better than the other algorithms in all the benchmarks. The frame length of PSA is significantly more than all algorithms, up to 25-40%, and PSA produces the maximal throughput. Except on benchmark III, the throughput of PSA is more than BSC by up to 37.5%. In benchmarks III, the BSC utilizes the channel better than PSA by up to 12.64% because the frame length of BSC is less than PSA by up to 22.02%.

## 5. CONCLUSION

The packet scheduling algorithm is to schedule packet in network layer and higher to reduce packet congestion in MAC layer and to reduce the packet collision and end-to-end delay; better packet delivery ratio is a by product. This algorithm is based on a greedy technique that is simple and easily implemented in a sensor node.

This paper measured the performance of the PSA with mathematical results in term of frame length, throughput, average delay, and channel utilization. The PSA is compared to previous works with network benchmarks. Our algorithm produces the highest throughput and utilizes the channel better than other algorithms. The PSA limitation is that the average delay is more than other algorithms. If we consider mathematical results only, it can not be concluded that any algorithm is suitable for wireless sensor networks. The PSA should be simulated and implemented on network simulation in order to determine performance in network perspective and we hop to publish the results soon.

## ACKNOWLEDGEMENTS

Authors thank Teacher Development Scholarship for the PhD program, Walailak University, Thailand which funded this research and Robert Elz for English proofing

## Authors

Chaiyut Jandaeng is a lecturer at the School of Informatics of Walailuk University. He completed the B.Sc and M.Sc. in Computer Science, PSU. He is a Ph.D candidate in Computer Engineering at Prince of Songkla University, Thailand. He was a teacher assistance and guest lecturer in information technology after completing his masters, before beginning the doctoral research. Areas of interest are computer networks security, and algorithm and programming technique in resource constrained devices.

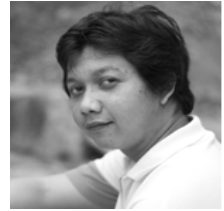

Wannarat Suntiamorntut received the Ph.D Degree from the University of Manchester, UK. She was a research assistant at the School of Computer Science, the University of Manchester for 3 years, while she was completing the Ph.D. program. Since 1999, she has been a lecturer at the computer engineering department at Prince of Songkla University, Hat Yai Thailand and is now assistant professor. In 2008, she became the Director of Collaborative Research Unit in Wireless Sensor Network (CRU-WSN), a joint collaboration between National Electronics and Computer Technology Center and Prince of Songkla University.

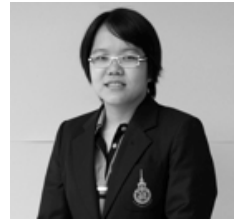

Nittida Elz  received her B.Sc. degree in Mathematics from Prince of Songkla University, Thailand. She then received her M.Sc. degrees in Computer Science from Chulalongkorn University, Thailand and The University of Melbourne, Australia, followed by a Ph.D. degree in Computer Engineering from La Trobe University, Melbourne, Australia. Her research interest is network management and security.

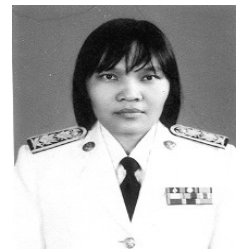